\documentclass[a4paper,fleqn,usenatbib]{mnras}
\usepackage[utf8]{inputenc}
\pdfoutput=1
\usepackage[T1]{fontenc}
\usepackage{ae,aecompl}
\usepackage{graphicx}
\usepackage{amsmath}
\usepackage{amssymb}
\usepackage{float}
\def\kms{\,km\,s$^{-1}$}

\title[Methanol variability in G111.256$-$0.770]{A highly variable methanol maser in G111.256$-$0.770}

\author[M. Durjasz et al.]{
M. Durjasz \thanks{E-mail: md@astro.umk.pl},
M. Szymczak
and M. Olech
\\
Centre for Astronomy, Faculty of Physics, Astronomy and Informatics, Nicolaus Copernicus University,\\
Grudziadzka 5, PL-87100 Torun, Poland
}

\date{Accepted XXX. Received YYY; in original form ZZZ}

\pubyear{2019}

\begin{document}
\label{firstpage}
\pagerange{\pageref{firstpage}--\pageref{lastpage}}
\maketitle

\begin{abstract}
G111.256$-$0.770 is a high-mass young stellar object associated with a weak 6.7\,GHz methanol maser showing strong variability.
We present results of a multi-epoch monitoring program of the target, conducted with the Torun 32\,m telescope for more than a decade.
We found that the isotropic maser luminosity varied by a factor 16 on a timescale of 5$-$6\,yr and individual features showed
small amplitude short-lived ($\sim$0.2\,yr) bursts superimposed on higher amplitude slow (>5\,yr) variations. 
The maser integrated flux density appears to be correlated with the near-infrared flux observed with the (NEO)WISE, supporting radiative pumping of the maser line.
\end{abstract}

\section{Introduction}
Methanol masers are one of the earliest observable signatures of high-mass young stellar objects (HMYSOs),
lasting $\sim10^4$\,yr \citep{vanderwalt2005} before or during the formation of ultra-compact H{\small II} regions
(\citealt{walsh1998}; \citealt{ellingsen2006}; \citealt{urquhart2013}; \citealt{devilliers2015}).
It is thought that they predominantly arise in the circumstellar discs or outflows (e.g. \citealt{moscadelli2016};
\citealt{sanna2017}) and one of the strongest transition at 6.7\,GHz probes the molecular gas of density
$10^4-10^8$\,cm$^{-3}$ and temperature below 150\,K \citep*{cragg2005}.

Recent studies have revealed the maser lines as sensitive indicators of sudden changes in the pumping conditions
in the environments of two HMYSOs likely triggered by accretion bursts \citep{caratti2017,moscadelli2017,hunter2018}.
Changes in the stellar luminosity due to accretion episodes appear to affect all maser features
and the overall maser structure at 6.7\,GHz in the HMYSO S255IR-IRS3 \citep{moscadelli2017} while turbulence in
the molecular clouds may produce short-lived (a week to a few months) random flares not correlated across different features
\citep*{goedhart2004}. When the maser beaming effect in an assembly of clouds of arbitrary geometry is considered then
quasi-periodic variations of the maser intensity may result from the rotation of maser clouds across the line of sight \citep*{gray2018}.
Long-lived (several months to a few years) bursts may be caused by outflows or shock waves passing through the masing region
(\citealt{goedhart2004}; \citealt*{dodson2004}). In the paper we report the results a long term monitoring of G111.256$-$0.770 (hereafter G111)
at 6.7\,GHz methanol transition which can shed more light on the processes affecting the masers.

The target is one of the most variable 6.7\,GHz masers in the sample studied by \cite{szymczak2018}; the relative amplitude
of individual features was up to 5.5 on timescales of between two months and a few years. The maser emission of isotropic luminosity
$8.5\times10^{-8}L_{\odot}$ \citep{wu2010} spatially coincides with the 22\,GHz H$_2$O masers and the radio continuum source of spectral
index 0.64 between 1.3 and 3.6\,cm wavelength is consistent with free-free emission from a thermal jet or with a partially thick
H{\small{II}} region \citep{trinidad2006}. Observations of outflow tracers revealed both redshifted and blueshifted components that largely
overlap on a scale lower than 10$-$15\arcsec\, suggesting an outflow almost completely along the line of sight \citep{wu2010,sanchez2013}
while the H$_2$ line emission knots and the radio continuum elongation delineate a collimated jet of size 5$-$7\arcsec\, at a position angle $-$44\degr\,\citep{massi2018}.
Most of the water maser components are blueshifted with respect to the systemic velocity of $-$44.5\,\kms\, (\citealt{sanchez2013}) and come
from a region of size up to 800$-$1000\,au \citep{goddi2005,moscadelli2016} for a trigonometric distance of 3.34\,kpc (\citealt{choi2014}).
Their proper motions have orientations both parallel and transverse to the direction of the jet \citep{moscadelli2016} suggesting
a low angle between the jet/outflow and the line of sight.

The present paper provides evidence that the 6.7\,GHz methanol maser from G111 is significantly variable on a decade timescale.

\section{Observations}
For the analysis presented in the paper we used published (\citealt{szymczak2018}) and unpublished archival
6.7\,GHz maser data and new observations, all of which were obtained with the Torun 32\,m radio telescope.
The new observations were carried out from February 2013 to October 2018.
The beam full width at half maximum of the antenna at 6.7\,GHz was 5.8\,arcmin and rms pointing error
was $\sim$25\,arcsec but it was reduced to 10\,arcsec from mid-2016 (\citealt{lew2018}).
The system temperature was between 25 and 40\,K. The data were dual polarization taken in frequency switching
mode using a 1\,MHz shift.  We used an autocorrelation spectrometer to acquire spectra with a  resolution
of 0.09\,\kms and the typical rms noise level of 0.35 Jy before 2015 May and 0.25 Jy afterwards.
The system was calibrated continuously against a noise diode of known constant temperature and this calibration
was daily checked by observing the non-variable maser source G32.744$-$0.076 and regular observations of the continuum
source 3C123 (\citealt{szymczak2018}). The resulting accuracy of the absolute flux density was better than 10 per cent.

To quantify the observed variability we used variability coefficients. A Gaussian function profile
fitting has also been performed. We estimated time scales with the structure function analysis and minimum-maximum method.
A more detailed description of the methods used is presented in Appendix A.

\section{Results} \label{sec:char}

\begin{figure}
\centering
\includegraphics[width=0.45\textwidth]{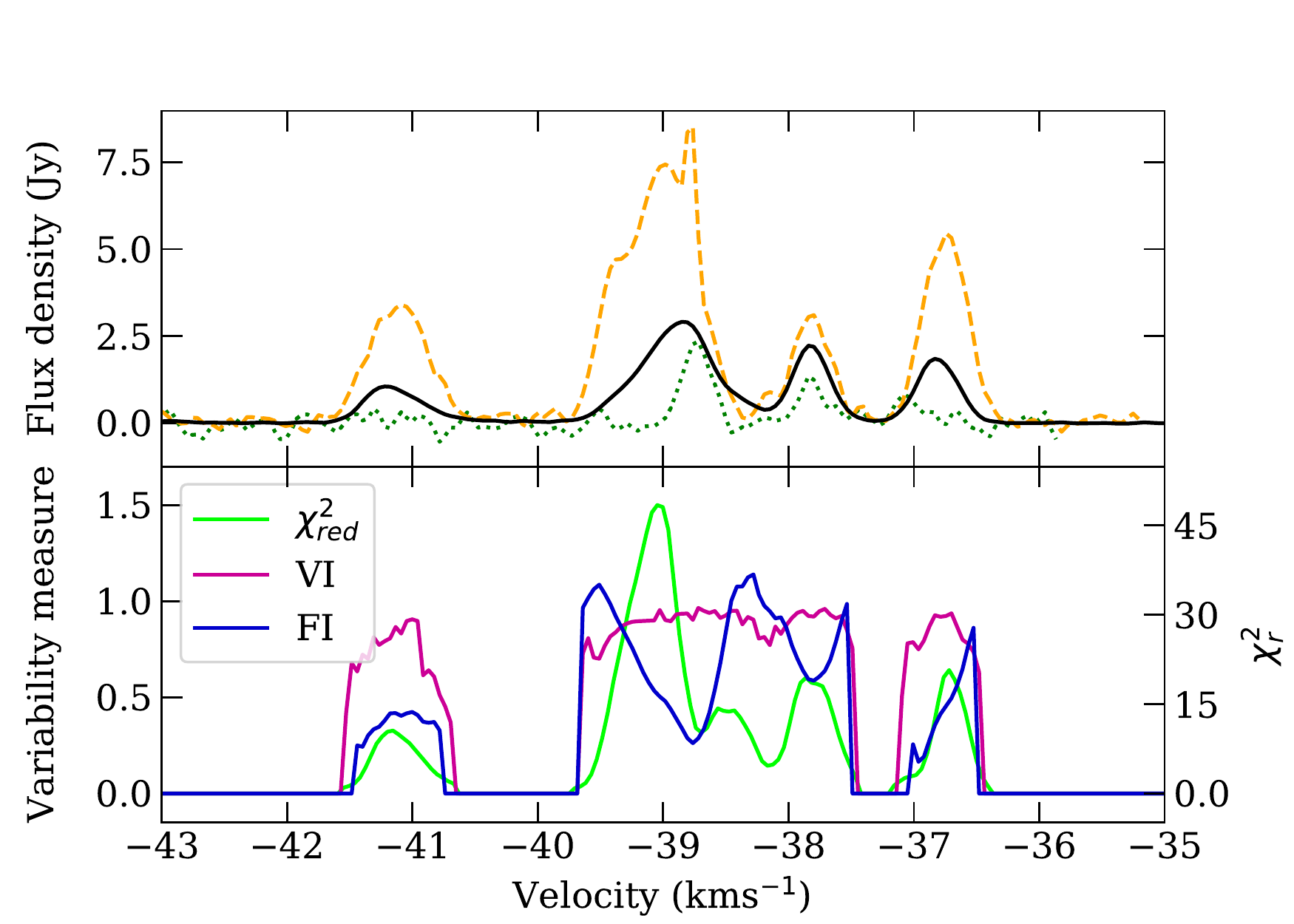}
\caption[Average spectrum of 6.7\,GHz methanol line]{{\it Top:} 6.7 GHz maser spectra of G111. The average (solid), high (dashed)
         and low (dotted) emission levels are shown. {\it Bottom:} Plots of variability index (VI), fluctuation index (FI) and reduced $\chi^2$.
\label{fig:aver-varind}}
\end{figure}

The average spectrum for the time-span from MJD 54583 to 58418 (685 observations) together with the spectra at high- and low-emission levels and plots of three variability measures are shown in Figure~\ref{fig:aver-varind}. The variability index ($VI$) and  fluctuation index ($FI$) quantify the flux amplitude change in variable sources \citep*{aller} and are described in Appendix A. All four persistent features displayed high variability over the whole observing period (Table~\ref{tab:var_ind_long}).

There are significant differences between $VI$ and $FI$ profiles which result from their properties;
$FI$ is sensitive to changes in the flux density with low signal-to-noise ratio and its value is highest in the wings of the features, whereas
$VI$ better depicts high amplitude variations.

\begin{figure*}
\centering
\includegraphics[width=0.85\textwidth]{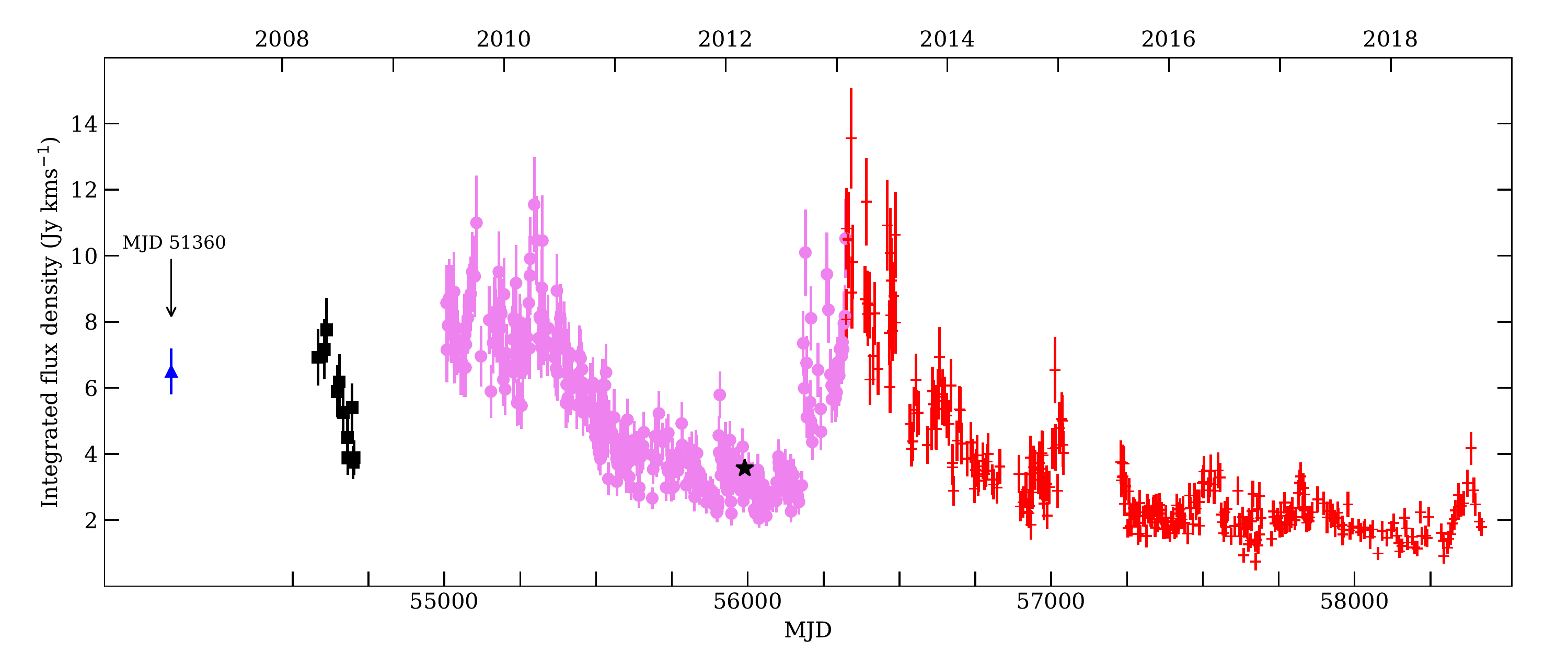}
\caption[Integrated CH$_3$OH spectrum]{Time series for the velocity-integrated flux density of the 6.7\,GHz methanol maser line.
The black squares and magenta circles denote the archival Torun 32\,m data and from \cite{szymczak2018}, respectively.
The blue triangle refers to the first detection (\citealt{szymczak2000}) and the black star marks the VLA observation at MJD 55990 (\citealt{hu2016}).
\label{fig:int_67}}
\end{figure*}

The light curve of the velocity-integrated flux density is presented in Figure~\ref{fig:int_67} where the new and archival observations
are complemented with the discovery detection (\citealt{szymczak2000}) and a single VLA observation (\citealt{hu2016}).
The emission peaked around MJD 55296 and 56340 and the second outburst has a FHWM of $\sim$256\,d and the rising phase a factor of two shorter than the declining phase.
In periods of low activity the emission linearly decreased with a rate of $-$0.17$\pm$0.02\,Jy\,km\,s$^{-1}$yr$^{-1}$.
The total flux density ranged from 0.9 to 14.5\,Jy\,km\,s$^{-1}$ and the average value was 4.4$\pm$0.1\,Jy\,km\,s$^{-1}$.
For the adopted distance of 3.34\,kpc the isotropic maser luminosity reached a peak of
1.0$\pm$0.2$\times10^{-6}L_{\odot}$ around MJD 56340 and declined to 8.9$\pm$1.9$\times10^{-8}L_{\odot}$ at the end of the monitoring period.
The median luminosity equals to 2.8$\times10^{-7}L_{\odot}$ is a factor of 3.2 lower than that reported for the sample observed in a high sensitivity,
untargeted survey with the Arecibo telescope (\citealt{pandian2009}). This confirms that the source belongs to a population of weak methanol sources
(\citealt{wu2010}).

\begin{table*}
\begin{center}
\caption{Variability measures of the 6.7\,GHz features. Variability index ($VI$), fluctuation index ($FI$) and $\chi^2_\mathrm{r}$ are
         for the mean velocity of feature while the average values $<VI>$, $<FI>$ and $<\chi^2_\mathrm{r}>$ and their standard errors are calculated for
         the given velocity range ($\Delta V$).
\label{tab:var_ind_long}}
\begin{tabular}{cccccccc}
\hline
$V$ & $VI$ & $FI$ & $\chi^2_\mathrm{r}$ & $\Delta V$ &  $<VI>$ & $<FI>$ & $<\chi^2_\mathrm{r}>$\\
(km\,s$^{-1}$) &   &      & &(km\,s$^{-1}$) &  & \\
\hline
$-$41.22  & 0.79 & 0.38 &  9.68 & $-$41.57;$-$40.65 & 0.68(0.18) & 0.32(0.10) & 5.44(3.15)\\
$-$38.85  & 0.93 & 0.34 & 24.25 & $-$39.68;$-$38.19 & 0.88(0.07) & 0.74(0.29) & 20.39(14.31)\\
$-$37.84  & 0.92 & 0.59 & 17.78 & $-$38.15;$-$37.40 & 0.85(0.14) & 0.79(0.15) & 11.67(6.31)\\
$-$36.83  & 0.93 & 0.36 & 10.97 & $-$37.18;$-$36.44 & 0.77(0.16) & 0.53(0.24) & 8.52(6.08)\\
\hline
\end{tabular}
\end{center}
\end{table*}

\begin{figure*}
\centering
\includegraphics[width=0.85\textwidth, angle=0]{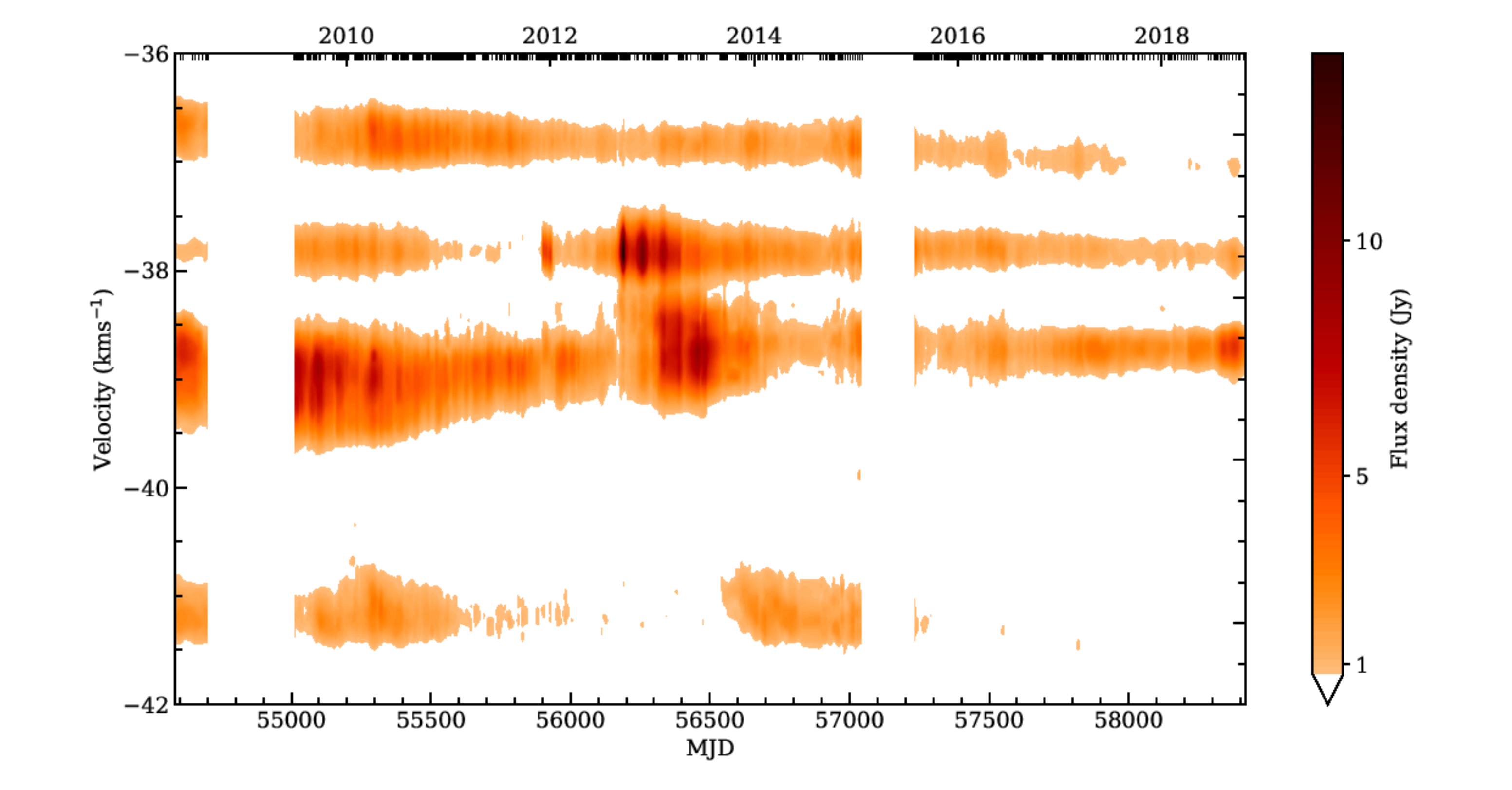}
\caption{False-colour image of the 6.7\,GHz maser flux density versus velocity and time. Velocities are measured in the frame of the local standard of rest. Individual observation dates are indicated by the vertical bars below the top horizontal axis. The two longest intervals with no data are blanked.
\label{fig:dyna68}}
\end{figure*}

The dynamic spectrum in Figure~\ref{fig:dyna68} visualizes the bulk variability and the bursting variability of 6.7\,GHz maser emission. It was created
using linear interpolation between consecutive 32\,m dish spectra and the emission above $3\sigma$ level is shown. There are four spectral features
in the spectrum and all of them display complex and high variability (Table~\ref{tab:var_ind_long}).

\begin{figure}
\centering
\includegraphics[width=0.45\textwidth]{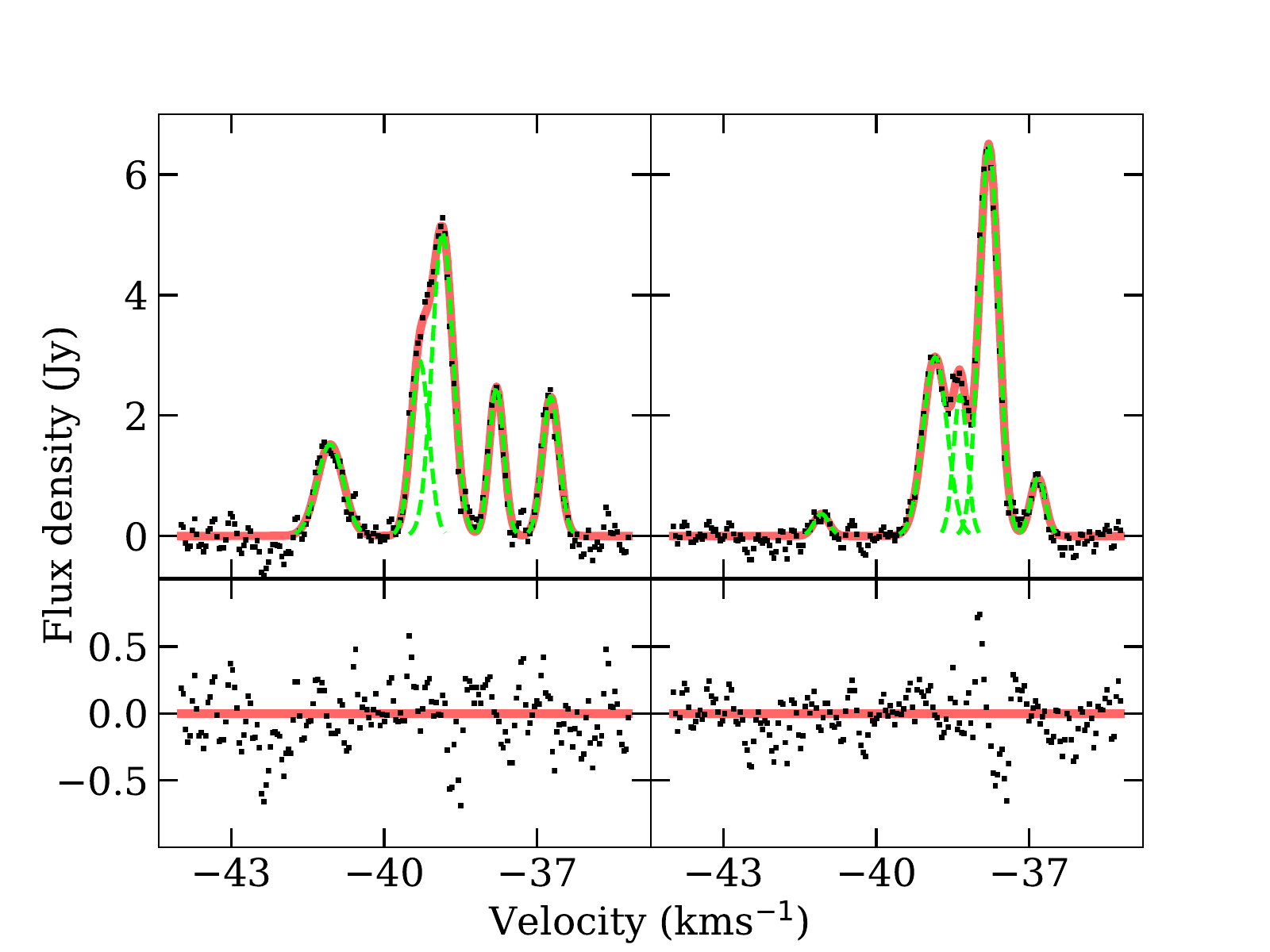}
\caption[New emission graph]{Examples of Gaussian profile fits to the spectra taken at MJD 55203 (left) and MJD 56276 (right). \label{fig:exfit}}
\end{figure}

\subsection{Specific variability}
In the following we deal with specific aspects of variability of the source.

\begin{figure}
\centering
\includegraphics[width=0.475\textwidth]{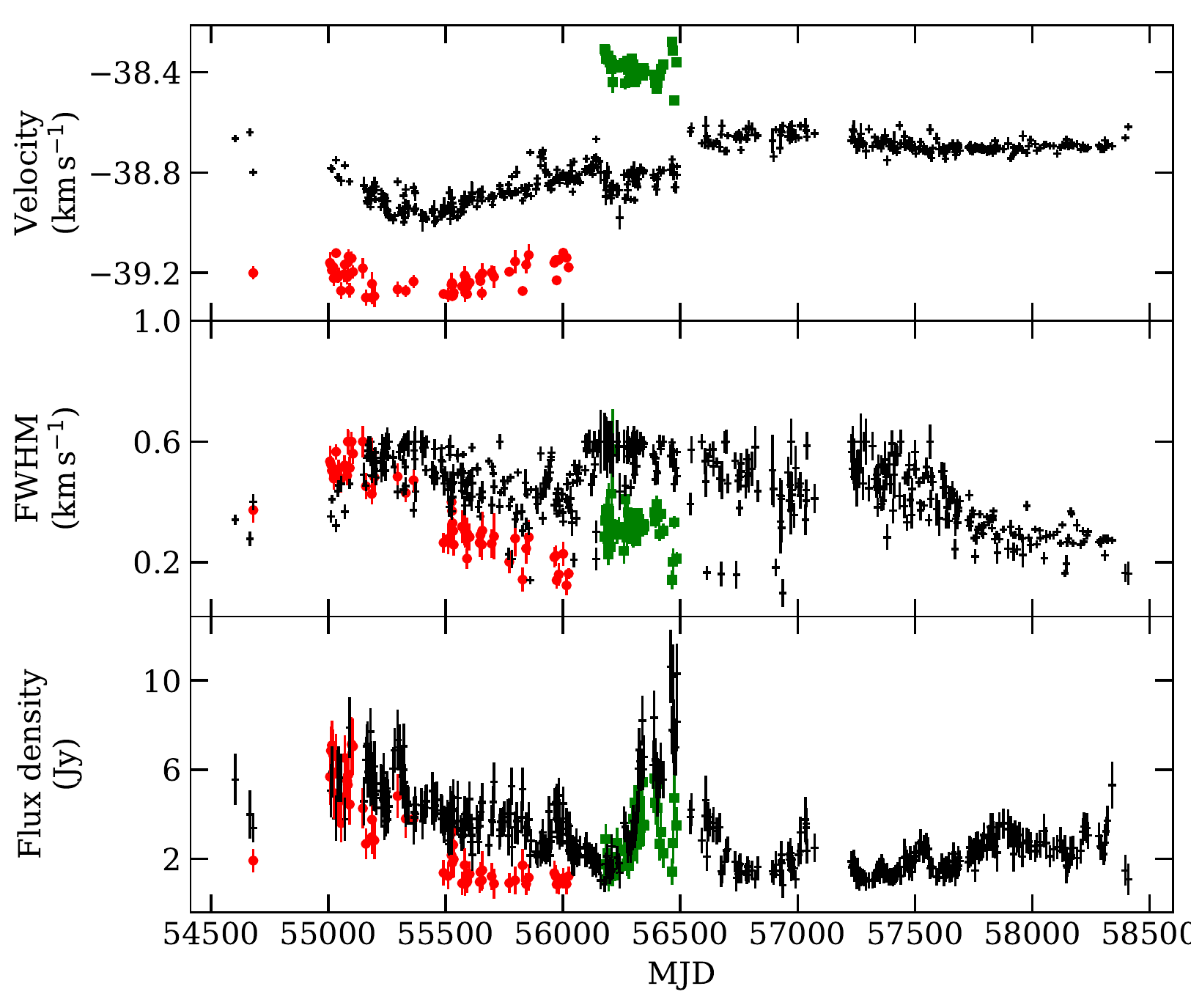}
\caption[Vlsr, fwhm and peak flux for $-$38.7\,km\,s$^{-1}$]{Peak velocity, line width at half maximum (FWHM) and peak flux density for Gaussian components of the emission near $-$38.7\kms. The red circles and green squares denote maser components appeared at slightly different velocities. \label{fig:amp_vlsr_387}}
\end{figure}

The Gaussian analysis of profiles revealed
a strong blending effect in some velocity ranges at different epochs (Fig.~\ref{fig:exfit}). For instance the emission of middle velocity features from $-$39.6 to $-$38.3\kms~ is composed of two Gaussian components at two time intervals of MJD 54680$-$56024 and MJD 56177$-$56485 (Figs.~\ref{fig:dyna68} and \ref{fig:amp_vlsr_387}). In the first and second time-spans the emission peaked at $-$39.15 and $-$38.85\kms\, and at $-$38.79 and $-$38.41\kms\,, respectively. Changes in the flux density and line width of the bursting features near $-$39.15 and $-$38.41\kms\, generally followed that for the persistent emission with a  mean velocity of $-$38.8\kms. From MJD 55418 to 57011 this persistent emission showed a drift in velocity of 0.071\kms~yr$^{-1}$.

\begin{figure}
\centering
\includegraphics[width=0.475\textwidth]{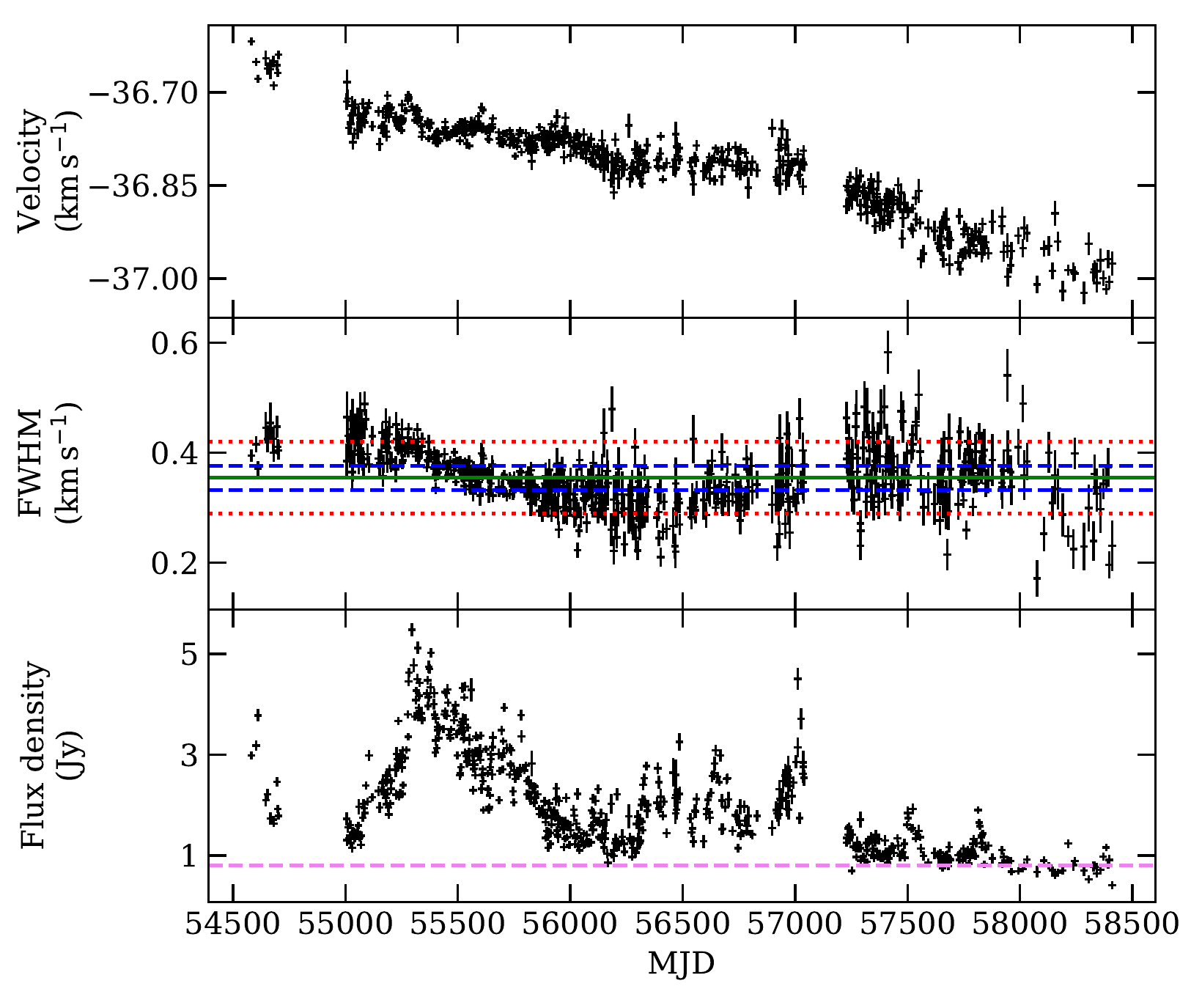}
\caption[Vlsr, fwhm and peak flux for $-$37\,km\,s$^{-1}$]{Same as in Fig.~\ref{fig:amp_vlsr_387} but for the emission near $-$36.8\kms. The green horizontal lines refer to the mean value of FWHM, the dashed violet line denotes the detection level of 0.8\,Jy. The blue (dashed) and red (dotted) lines mark 1$\sigma$ and 3$\sigma$ levels, respectively. \label{fig:amp_vlsr_37}}
\end{figure}

Figure~\ref{fig:amp_vlsr_37} illustrates the behaviour of the emission near $-$36.7\kms. The feature showed a long lasting ($\sim$700\,d) burst with a peak of $\sim$5\,Jy around MJD 55295 and several short ($<$200\,d) less visible bursts and it dropped below our sensitivity level at the end of monitoring. No significant variations in the FWHM were seen. A velocity drift of $-$0.019\kms\,yr$^{-1}$ occurred before MJD 57039 and changed to $-$0.043\kms\,yr$^{-1}$ afterwards.

\begin{figure}
\centering
\includegraphics[width=0.475\textwidth]{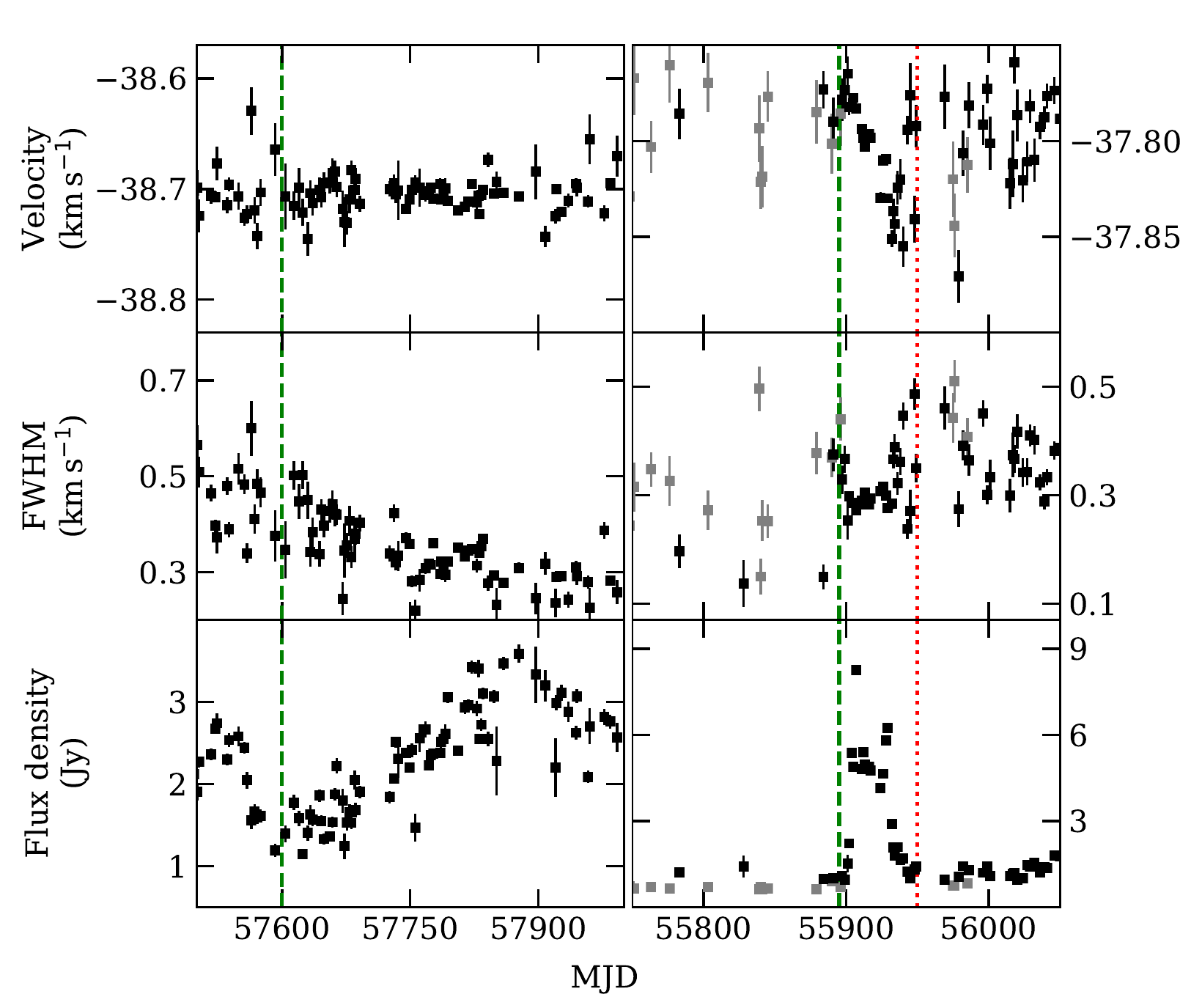}
\caption[Flares]{Same as in Fig.~\ref{fig:amp_vlsr_387} but for two exemplary bursts. Left and right panels refer to $-$38.73 and $-$37.88\kms\, features, respectively.
The green vertical lines denote the start time of bursts and the red vertical marks the end of burst. The grey symbols represent the measurements with the flux density below 3$\sigma$
level.
\label{fig:flares}}
\end{figure}

Figure \ref{fig:flares} depicts the variability of two features during the bursts. For the feature $-$38.73\kms\,
the rising phase of a burst is clearly seen; the flux density increased from $\sim$1.2 to $\sim$3.6\,Jy over $\sim$306\,d, the peak velocity remained stable but the FWHM decreased from 0.50 to 0.24\kms. This may be attributed to unsaturated amplification despite the fact that the canonical relationship between the intensity and linewidth \citep{goldreich1974} is not fulfilled. On the other hand the declining phase is poorly seen and there is no evidence for profile broadening after the maximum.

The emission near $-$37.88\,km\,s$^{-1}$ showed a short (49\,d) burst with rising and declining phases of 11 and 38\,d, respectively. It is striking that the line width was nearly constant while the peak velocity
drifted by 0.086\kms\, during a $\sim$31\,d
long burst. This phenomenon can appear when a factor triggering the burst propagates to nearby layers of similar gas temperature but slightly different radial
velocity. As the maser features are redshifted relative to the systemic velocity the drift may trace inflow motion.

We conclude, the variability of 6.7\,GHz maser in the source is characterised by short duration (2--5\,months) bursts at slightly different velocities superimposed
on long lasting (1.5$-$3\,yr) high relative amplitude (4$-$16) changes in the flux density showing velocity drifts (up to $\sim$0.1\kms\,yr$^{-1}$) for some features.

\subsection{Variability timescales and delays}
\begin{figure}
\centering
\includegraphics[width=0.475\textwidth]{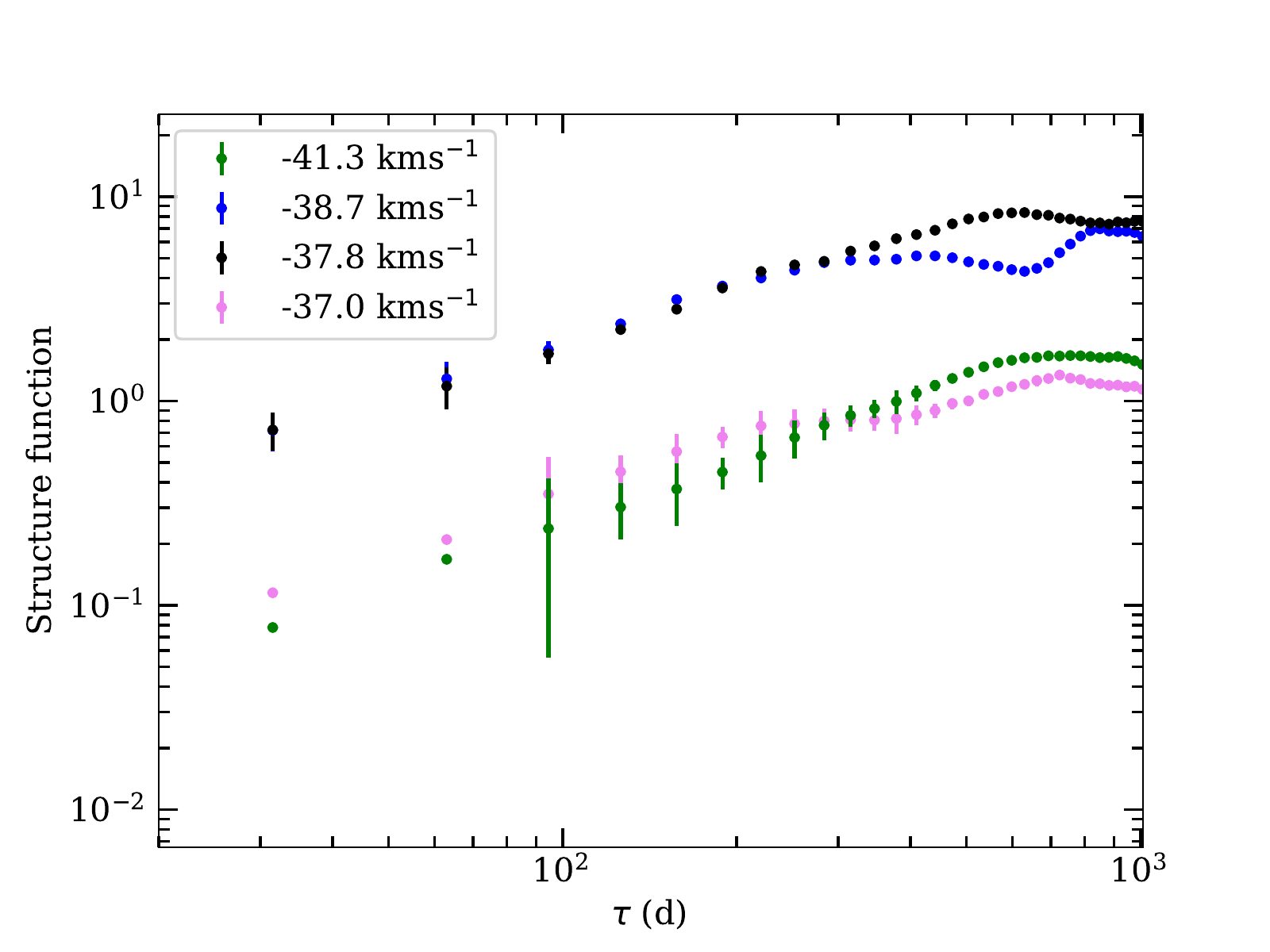}
\caption[Structure functions]{Structure functions of the persistent maser features.\label{fig:sf38p7}}
\end{figure}

In order to determine  a variability timescale we used the discrete structure function and minimum-maximum method (see Appendix A, \citealt{fuhrmann2008}).
The structure functions for the four persistent features are presented in Figure \ref{fig:sf38p7} and values of variability timescales obtained with both methods
are given in Table~\ref{tab:tvar}.
The timescales for the features $-$41.22 and $-$36.83\kms\, are 1.5 to 1.7\,yr. The uncertainties of these estimates are usually lower than 26~per~cent with
the exception of the $-$36.83\kms\, feature that shows only a monotonic decrease and $t_\mathrm{sf}$ could not be accurately determined.

\begin{table}
\begin{center}
\caption{Variability timescales for the spectral features. $t_\mathrm{var}$ and $t_\mathrm{sf}$ are the timescales
calculated with the minimum-maximum method and the structure function, respectively. \label{tab:tvar}}
\begin{tabular}{ccc}
\hline
Feature (\kms) & $t_\mathrm{var}$ (d) & $t_\mathrm{sf}$ (d)\\
\hline
\vspace{0.1cm}
$-$41.22 & 506.1 $\pm$ 130.9 & 605.5$^{+60.1}_{-69.9}$\\
\vspace{0.1cm}

$-$38.85 & 107.7 $\pm$ 10.7 & 274.9$^{+58.1}_{-82.5}$\\
\vspace{0.1cm}

$-$37.84 & 45.8 $\pm$ 5.1 & 533.3$^{+90.7}_{-118.2}$\\

$-$36.83  & 608.1 $\pm$ 104.9 & 663.6$^{+198.2}_{-323.7}$\\
\hline
\end{tabular}
\end{center}
\end{table}

The values of the variability timescales for features at $-$38.85 and $-$37.84\kms\, obtained by the two methods differ significantly (Table~\ref{tab:tvar}).
This is probably due to the fact that $t_\mathrm{var}$ strongly relies on the individual choice of the minimum and maximum measurements that make it
irrelevant for describing  the long lasting burst variability. We conclude that the variability timescales of the methanol masers in G111
range from 0.7 to 1.8\,yr.

\begin{figure}
\centering
\includegraphics[width=0.475\textwidth]{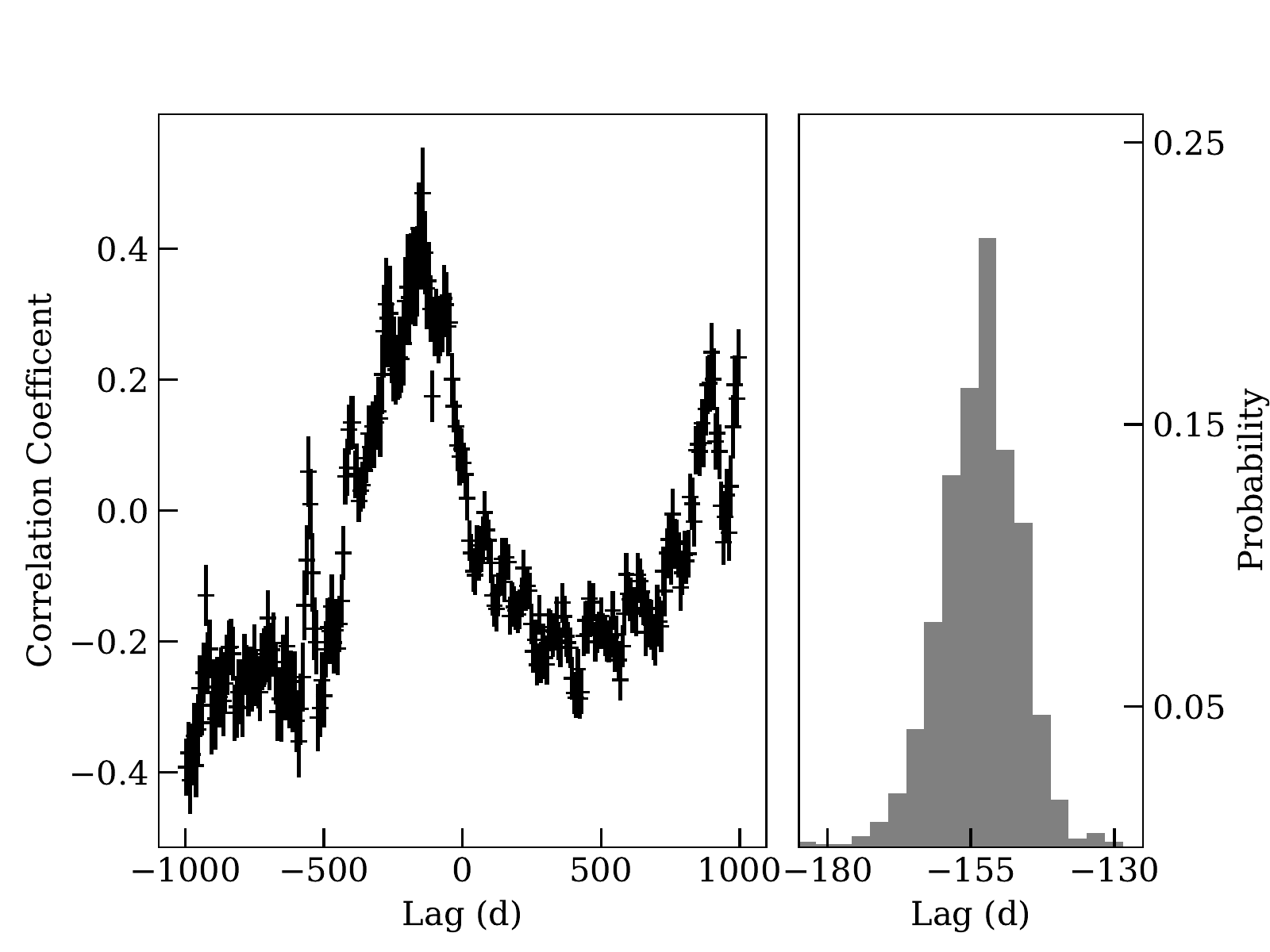}
\caption[Dcf and CCPD]{{\it Left.} Example of discrete correlation function between the light curves of features $-$38.85 and $-$37.84\kms. {\it Right.} Cross-correlation peak distribution
for the peak of the DCF. \label{fig:ccpdhis}}
\end{figure}

In order to search for time lags we computed the discrete cross-correlation function (DCF, \citealt{edelson1988}) between the three spectral features.
The maximum value was obtained with the centroid $\tau_\mathrm{c}$ of the DCF given by
$\tau_\mathrm{c} = {\sum_\mathrm{i} \tau_\mathrm{i} \mathrm{DCF}_\mathrm{i}}/{\sum_\mathrm{i} \mathrm{DCF_i}}$.
Following the method presented in detail by \cite{peterson1998} and \cite{fuhrmann2008} we performed Monte Carlo simulations to statistically estimate meaningful time
lags and their uncertainties. The influence of uneven sampling and the measurement errors were taken into account by using random subsets of
the two spectral features time series and by adding random Gaussian fluctuations constrained by the measurement errors.
For two pairs of the spectral features ($-$38.85\,km\,s$^{-1}$ vs $-$37.84\,km\,s$^{-1}$ and $-$38.85\,kms$^{-1}$ vs $-$36.83\,km\,s$^{-1}$) the centroid of the DCF maximum was computed 1000 times.
The DCF and cross-correlation peak distribution (CCPD) for one pair of features are shown in Figure \ref{fig:ccpdhis}. Table~\ref{tab:dcf} summarizes the results for two pairs of features.
The time delay of the peaks estimated between the two maser features ($-$38.85 and $-$37.83\kms) in the  timespan from $\sim$MJD 56200 to 56900 is 5 months.
The delay of the peak of the redshifted emission ($-$36.83\kms) has a too large uncertainty to be considered as reliable. A visual inspection of the light curves (Fig.~\ref{fig:dyna68})
implies that the burst of feature $-$37.84\kms\, around MJD 56200 was advanced by 147\,d relative to the burst of feature at $-$38.85\kms.
This crude estimation is in good agreement (within 20\,per~cent) with that presented in Table~\ref{tab:dcf}.

\begin{table}
\begin{center}
\caption{Time lags corresponding to the average centroids of the DCF peaks obtained from cross-correlation peak distributions. Uncertainty values correspond to 1$\sigma$ obtained from CCPDs. \label{tab:dcf}}
\begin{tabular}{cc}
\hline
Features (\kms)& $\tau$ (d) \\
\hline
$-$38.85 v.s. $-$37.84 & $-$153.7 $\pm$ 6.8\\
$-$38.85 v.s. $-$36.83 & 210.4 $\pm$ 80.3\\
\hline
\end{tabular}
\end{center}
\end{table}

\section{Discussion}

\begin{figure}
\centering
\includegraphics[width=0.45\textwidth]{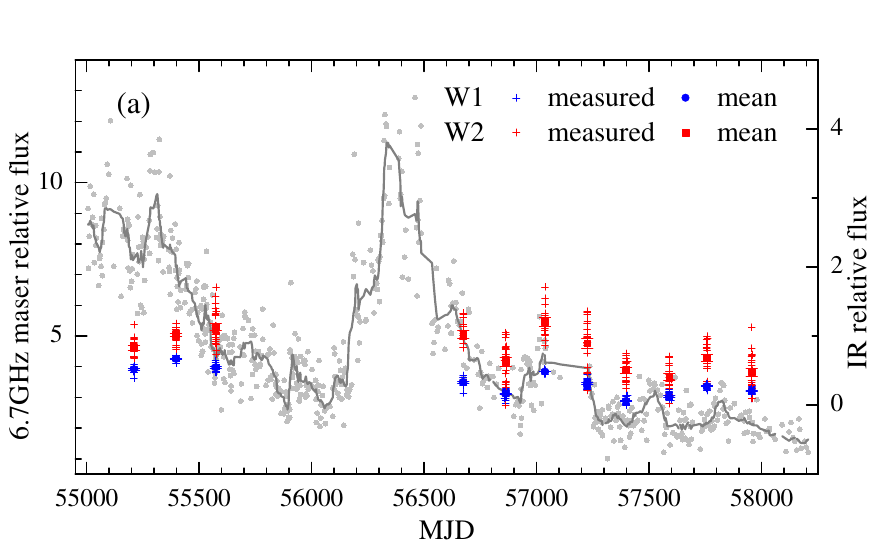}
\includegraphics[width=0.45\textwidth]{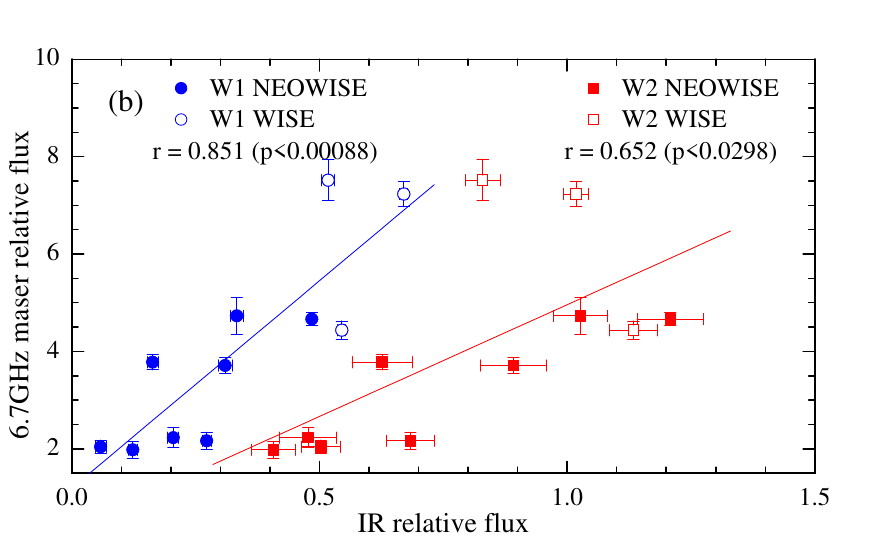}
\caption[Maser infrared relationship]{({\it a}) Part of the maser light curve from Fig.~\ref{fig:int_67} with the IR relative fluxes superimposed. The maser relative velocity-integrated flux density
        and its 5 point averaged values are marked by grey circles and grey line, respectively. IR data are from WISE and NEOWISE observations at bands W1 ($3.4\mu$m, blue crosses)
        and W2 ($4.6\mu$m, red crosses). The corresponding relative average IR fluxes are denoted by blue circles and red squares, respectively. ({\it b}) Relationships between the maser and
        IR relative fluxes. The solid lines denote the linear least-square fits to the data.
 \label{fig:mas-ir}}
\end{figure}

Our monitoring revealed that the methanol emission from G111 has experienced complex and erratic variations for the last decade. In the following we discuss the observed characteristics
which may shed light on the processes causing variability.

One of the important factors producing the maser variability can be changes in the pump rate. Since the 6.7\,GHz transition is thought to be pumped by infrared photons \citep{sobolev1997,cragg2005}
we examine if the maser emission is related to the near infrared radiation.
Data from the WISE and NEOWISE archives \citep{Wright2010,Mainzer2011} were retrieved for the target to construct the light curves at 3.4 and 4.6$\mu$m. In Figure~\ref{fig:mas-ir}a we show
an overlay of the relative IR light curves with the methanol maser light curve. The WISE and NEOWISE photometric observations were available for 11 sessions of length of
1 to 8\,d (median 1.8\,d) over time-span 2010$-$2017 exactly covering our monitoring period. The changes in the IR fluxes generally followed those observed
in the 6.7\,GHz integrated (over the whole spectrum) maser flux in the declining phase of the long lasting ($>$5\,yr) burst. Monthly averages of the integrated flux density
were compared with the mean IR flux at 11 epochs and Figure~\ref{fig:mas-ir}b shows a significant correlation between the relative IR and maser fluxes. Although the IR data are not available
during the maser burst around MJD 56400 and the trend in the IR relative brightness over the time-span $\sim$MJD 55200$-$55570 does not follow a decrease in maser intensity, this relationship may support
a scenario in which the 6.7 GHz maser is radiatively pumped and changes in the pump rate may cause the 6.7\,GHz maser variability as it was firmly demonstrated for maser
sources experiencing giant bursts \citep{caratti2017, hunter2018}. IR observations of sufficient cadence are required to confirm this relationship for other variable sources.

We detected variations in the centroid velocity of the two maser features. The velocity drift of feature $-$38.7\kms\, before $\sim$MJD 57000 can be easily recognized as due to blending effect
of two features showing temporal intensity variations because velocity jumps were clearly seen over certain timespans (Fig.~\ref{fig:amp_vlsr_387}). A similar effect may occur for the $-$36.8\kms\,
feature with  FWHM ranging from 0.30 to 0.45\kms\, over the monitoring period (Fig.~\ref{fig:amp_vlsr_37}) and the  slope of the drift changed to steeper after $\sim$MJD 57000. In this case we cannot
resolve blending components probably due to the low signal-to-noise ratio. The opposite sense of the velocity drifts reinforces the interpretation that the velocity of emission fluctuates rather
than drifting steadily. One can conclude that the velocity drifts are caused by slow (<10\,yr) variations in the flux density of a few features at very close velocities.

In G111 we sometimes observed correlated variations across two different features combined with short-lived (2$-$3\,months) flares restricted to one feature (Figs.~\ref{fig:dyna68}, \ref{fig:flares}).
If this variability timescale is on the order of the shock crossing time, which determines the lifetime of velocity coherence along the line-of-sight necessary for the maser amplification
then for a typical size of the methanol maser cloud of 4\,au \citep{bartkiewicz2016} the  shock velocity is larger than 70\kms. It very unlikely that the methanol maser would survive such
a shock. Random bursts indicate rather local changes in the maser conditions modified for instance by turbulence \citep{sobolev1998}; a dispersion of turbulence velocity of about 1\kms\, could
account for the observed short-lived bursts caused by changes in the velocity coherence.

\section{Summary}
We report new 6.7\,GHz methanol maser observations of G111 obtained since 2013 which extend the previously published light curve to $\sim$11\,yr.
The maser emission is characterised by short duration (2-3 months) bursts superimposed on long lasting ($>$5\,yr) variations with a relative amplitude
of 4 to 16.  The comparison of the maser integrated flux density with the  near infrared emission may
support the radiative pumping scheme of the maser line but infrared observations of sufficient cadence are required to draw  a firm conclusion.

\section*{Acknowledgements}
We thank the Torun CfA staff for assistance with the observations. We appreciate Eric  G\'erard for carefully reading the manuscript and providing critical comments.
This research has made use of the SIMBAD data base, operated at CDS (Strasbourg, France), as well as NASA's
Astrophysics Data System Bibliographic Services. This work has also made use of data products from the Wide-field Infrared Survey Explorer, which is a joint project of the University of California,
Los Angeles, and the Jet Propulsion Laboratory/California Institute of Technology, funded by the National Aeronautics and Space Administration
and from NEOWISE, which is a project of the Jet Propulsion Laboratory/California Institute of Technology, funded by the Planetary Science Division of the National Aeronautics and Space Administration.
The work was supported by the National Science Centre, Poland through grant 2016/21/B/ST9/01455.

\bibliography{librarian}{}
\bibliographystyle{mnras}

\appendix
\section{Analysis methods}
\subsection{Statistical variability measures}
We used the variability index as proposed by \cite{aller}

\begin{equation}
VI = \frac{ (S_{\mathrm{max}} - \sigma_{\mathrm{Smax}}) - (S_{\mathrm{min}} - \sigma_{\mathrm{Smin}}) }{ (S_{\mathrm{max}} - \sigma_{\mathrm{Smax}}) + (S_{\mathrm{min}} - \sigma_{\mathrm{Smin}})},
\label{eq:VI}
\end{equation}
where $S_{\mathrm{max}}$ and $S_{\mathrm{min}}$ refer to the maximum and minimum flux densities, respectively, while $\sigma_{\mathrm{Smax}}$ and
$\sigma_{\mathrm{Smin}}$ are the corresponding uncertainties of $S_{\mathrm{max}}$ and $S_{\mathrm{min}}$. This index is useful to estimate the amplitude
of the variability accounting for the measurement uncertainties and is well determined when variability is much greater than measurement errors.

The second variability measure used is the fluctuation index \citep{aller}

\begin{equation}
FI = \left[ \left(\frac{\sum S^{2}_{\mathrm{i}}w_\mathrm{i} - \overline{S}\sum S_\mathrm{i}w_\mathrm{i}}{N - 1} - 1 \right) \frac{N}{\sum w_\mathrm{i}} \right]^{0.5}\frac{1}{\overline{S}},
\label{eq:FI}
\end{equation}
where $N$ is the number of observations of the flux density $S_{\mathrm{i}}$ measured with error $\sigma_{\mathrm{i}}$, weight is $w_{\mathrm{i}}=\sigma_i^{-2}$ and $\overline{S}$ is
the average flux density. This index well estimates variability of features with low signal-to-noise ratio.

We also examined variability by computing the reduced value of  $\chi^2$.
\begin{equation}
\chi^2_{\mathrm{r}} = \frac{1}{N-1} \sum^{N}_{i=1} \left( \frac{S_\mathrm{i} - \overline{S}}{\sigma_\mathrm{i}} \right)^2
\label{eq:chi2}
\end{equation}

\subsection{Time scales of variability}
We calculated the discrete structure function
\begin{equation}
SF(\tau_{\mathrm{j}}) = \sum_{i=1}^{n}[S(t_\mathrm{i}) - S(t_\mathrm{i}+\tau_\mathrm{j})]^2
\label{eq:scf}
\end{equation}
for each data set $S(t_\mathrm{i})$ following the procedure described in \cite{heszen1987}.
Here, $n$ is the number of flux density measurements obtained at epochs $t_\mathrm{i}$ and $t_\mathrm{i}+\tau_\mathrm{j}$.
We binned the data in 31\,d intervals to get evenly sampled dataset. Two linear fits were performed to estimate the slope
($a\tau^{\beta}$) of the $SF$ before the first maximum and the mean value of $SF$ after the first maximum which estimates
the saturation level $\rho_0$. The timescale of variability was then determined from formula
\begin{equation}
t_\mathrm{sf} = \left(\frac{\rho_0}{a} \right)^\frac{1}{\beta}
\label{eq:tsf}
\end{equation}

\bsp
\label{lastpage}
\end{document}